\documentstyle[12pt]{article}
\newcommand{\beq}{\begin{equation}}
\newcommand{\eeq}{\end{equation}}

\def\half{{\textstyle{1\over2}}}

\def\p1half{{\textstyle{{{p+1}\over{2}}}}}

\def\23phalf{{\textstyle{{{23-p}\over{2}}}}}

\begin{document}
\thispagestyle{empty}
\begin{titlepage}

\bigskip
\hskip 5.2in{\vbox{\baselineskip12pt
\hbox{hep-th/0105110}}}

\bigskip\bigskip\bigskip\bigskip
\centerline{\large\bf
Finite Temperature Bosonic Closed Strings:}

\medskip
\centerline{\large\bf
Thermal Duality and the Kosterlitz-Thouless Transition}

\bigskip\bigskip
\bigskip\bigskip
\centerline{\bf Shyamoli Chaudhuri
\footnote{Current Address: 1312 Oak Dr, Blacksburg, VA 24060. Email: shyamolic@yahoo.com}
}
\centerline{214 North Allegheny St.}
\centerline{Bellefonte, PA 16823}
\date{\today}

\bigskip\bigskip
\begin{abstract}
We elucidate the properties of a gas of free closed bosonic strings in thermal 
equilibrium. Our starting point is the intensive generating functional of 
connected one-loop closed vacuum string graphs given by the Polyakov
path integral. Invariance of the path integral under modular transformations 
gives a thermal duality invariant
expression for the free energy of free closed strings at finite temperature.
The free bosonic string gas exhibits a self-dual Kosterlitz-Thouless phase 
transition. The thermodynamic potentials of the gas of free 
bosonic closed strings are 
shown to exhibit an infinite hierarchy of thermal self-duality relations.
Note Added (Sep 2005).
\end{abstract}
\noindent

\end{titlepage}

\section{Introduction}

A reliable and self-consistent analysis of the thermodynamics of free
strings is an essential step prior to the formulation of a microscopic
and fully nonperturbative framework at finite string coupling that can
cope with the physics of black holes, the formation of spacetime horizons 
and spacetime singularities, and the earliest evolutionary stages of 
the Universe 
at Planckian scales. Surprisingly, although the finite temperature 
behavior of an ensemble of free strings has been extensively studied 
since the early days of string theory
\cite{huangw,bowick,poltorus,longs,tan,sathia,aw} there are few 
conclusive results, even in this simplest of cases. 
A dominant theme in early works is that 
of a limiting temperature beyond which a gas of 
free strings is thought to become unstable with an exponentially diverging 
Helmholtz free energy. The notion of the Hagedorn phase transition, 
in fact, predates the world-sheet formalism of string theory 
and was originally proposed within the context of phenomenological dual 
reggeon models for the hadronic bound state spectrum \cite{hagedorn,carlitz}. 
The motivation for the suggestion comes from the fact that the asymptotic 
density of states function for a zero temperature gas of free strings is
known to exhibit exponential growth \cite{huangw,hr}. A simple 
back-of-the-envelope estimate reviewed in the appendix 
indicates that, at least for particle-like thermodynamic ensembles, an
exponential growth in the density of states function implies an 
exponentially diverging free energy beyond some characteristic temperature
$T_H$. This abrupt change in the behavior of the free energy at $T_H$
has been interpreted as a thermal phase transition \cite{carlitz} and 
is known in the literature as the Hagedorn phase transition. 

\vskip 0.1in
This intuition relies, however, on the assumption that the thermal behavior
of a free string gas is well-approximated by that of infinitely many
point particle modes with a level spectrum of integer-spaced Planck-scale 
masses. We will show in this paper, and even more conclusively in \cite{fermi}, 
that the lore of a Hagedorn phase transition in perturbative string theory 
is untrue. As already emphasized by Polchinski in \cite{poltorus},
the free energy of a gas of free strings is not the sum of the free 
energies of infinitely many point-particle field theories with 
Planck scale masses.
The free energy at one-loop 
order in string theory is expressed as an integral over the conformally 
inequivalent classes of a torus 
parameterized by a complex worldsheet modulus $\tau$ 
\cite{poltorus,polchinskibook}, or over the fundamental domain
of the modular group. The bosonic string theory has a zero temperature 
tachyon, as a consequence of which the Polyakov path
integral exhibits an infrared divergence. However, modular 
invariance permits one to formally map the IR divergence into what
would appear to be a UV divergence \cite{mclain,tan,polchinskibook}
by mapping the fundamental domain to a rectangular strip in the
$\tau$ plane. This mapping obscures the physical (infrared) 
origin of the tachyonic divergence. In the tachyon-free heterotic 
string gas, it will be straightforward to demonstrate that the 
original worldsheet representation as an integral over the fundamental 
domain is perfectly adequate as a path integral expression for the 
vacuum functional at finite temperature. The path integral expression 
for the free energy of the heterotic string gas has neither ultraviolet 
nor infrared divergences at all temperatures starting from zero \cite{fermi}. 
Thus, there is no signal of a Hagedorn transition in the oneloop
free energy. In the type I open and closed string gas, a similar result 
follows from the absence of both dilaton tadpoles and tachyons 
in the open string spectrum \cite{us,fermi}. 
This paper serves as a warm-up to these
latter works with a study of the pedagogical case 
of the closed bosonic string gas. 

\vskip 0.1in
Of great interest, we will find concrete evidence for the
occurence of a Kosterlitz-Thouless 
continuous phase transition responsible for the formation of a 
\lq\lq long string phase" in 
the high temperature gas of free open and closed strings 
\cite{longs,polchinskibook,us,fermi}. 
In the bosonic and heterotic closed string gases the 
phase transition 
instead takes the form of a self-duality transition.
We will show that the thermodynamic potentials characterizing
the duality transition in free string gases can be computed 
in each case in transparent detail \cite{fermi}. 
The thermodynamic potentials satisfy an infinite 
hierarchy of duality relations at criticality.
 
\vskip 0.1in
Our starting point is the Polyakov path integral representation for
the one-loop contribution to the vacuum energy density in string theory
\cite{polyakov,poltorus}. Denoted as $W(\beta)$, it gives the
sum over connected random surfaces with the topology of a torus in the
Euclidean spacetime $R^{25}$$\times$(a compact one-dimensional target 
space whose volume corresponds to the inverse temperature, $\beta$).
The generating functional for connected one-loop vacuum string graphs
is an {\em intensive} thermodynamic variable on which we impose Euclidean
T-duality relations. For the self-dual closed bosonic string, this implies
invariance of the vacuum string functional: $W(\beta)$ $\equiv$ ${\rm ln}$ $Z(\beta)$, under 
thermal duality transformations: $\beta$ $\to$ $\beta_C^2/\beta$.
In addition, $W(\beta)$ is required to be invariant under modular transformations
as a consequence of its definition as a reparametrization invariant 
sum over Riemann surfaces \cite{polyakov,poltorus}.
We cannot of course achieve a tachyon-free thermal spectrum for the
bosonic string but our computation gives a simple illustration of the 
general framework of string thermodynamics. 
The reader will find that it is straightforward to infer the corresponding results 
in the tachyon-free self-dual heterotic string ensemble \cite{fermi}. 

\vskip 0.1in
The path integral representation of $W$ will
lead directly to the Helmholtz function: $F(\beta)$ $=$ $-T$ ${\rm ln }Z(\beta)$, also
known as the Helmholtz free energy, and the finite temperature effective 
potential: $\rho(\beta)$ $=$ $-{{T}\over{V}}$ ${\rm ln}Z(\beta)$. 
The internal energy, $U$, pressure, $P$, entropy, $S$, enthalpy, $H$,
and specific heat, $C_V$, of the free closed 
bosonic string gas will be derived by simply
taking partial derivatives with respect to temperature, or spatial 
volume, of the modular and thermal duality invariant vacuum string 
functional, ${\rm ln}$ $Z(\beta)$. We will
find that the pressure of the free bosonic string gas vanishes in
the absence of string interactions and that the enthalpy consequently
equals the internal energy. As a result the Gibbs free energy of
the free string gas, $G$, is found to coincide with the Helmholtz 
free energy, $F$:
$G$$=$$H$$-$$TS$$=$$U$$+$$PV$$-$$TS$$=$$U$$-$$TS$$=$$F$.
The thermodynamic potentials in general are found to satisfy an infinite
hierarchy of thermal duality relations derived in section 4. 
Previous analyses of bosonic closed string thermodynamics in the
canonical framework either neglect, or explicitly violate, thermal 
duality \cite{bowick,poltorus,mclain,longs,tan,sathia,aw}. An 
exception is the 
thermal duality invariant closed bosonic string spectrum described 
in the recent text \cite{polchinskibook}. 
We should note that thermal duality follows naturally in a closed string
theory as a consequence of worldsheet modular invariance. 

\vskip 0.1in
The plan of this paper is as follows. In section 2, we begin with a 
brief review of some aspects of finite temperature field theory in
the canonical formalism, giving a summary of the main thermodynamic 
potentials and thermodynamic identities that will be of interest in
our discussion. In section 3.1, we obtain both the normalization and 
phase of the generating functional of one-loop vacuum string 
graphs at finite temperature. The normalization is obtained following 
the method developed in \cite{poltorus,polchinskibook}. 
We show that the temperature dependent phases 
in the string path integral are unambiguously determined 
by modular invariance or, equivalently, by thermal duality.

\vskip 0.1in
We begin by noting an ambiguity in the Euclidean time prescription 
which becomes apparent 
when we try to apply it to finite 
temperature string theory. Surprisingly, this has not been pointed out 
prior to the appearance of our work \cite{us}.
The topology of a compact one-dimensional target space is either that
of a circle, $S^1$, with inverse temperature $\beta$ identified as the 
circumference, $2\pi r_{\rm circ}$, or, that of an orbifold, $S^1/Z_2$, 
where the $Z_2$ acts on the circle as a reflection: $X^0$$\to$$-X^0$,
with fixed points at $X^0$$=$$0$ and $X^0$$=$$\pi r_{\rm circ}$,
and $\beta$ identified as the interval length, $\pi r_{\rm circ}$. Being 
extended objects, free strings can explore the global topology of the 
space in which they live. Consequently, both the free string mass 
spectrum, and the free energy of the free string gas, are expected to 
distinguish between the two possible topologies for Euclidean time. 
Which gives the correct answer? 
For finite temperature field theories, as mentioned already in 
\cite{us,fermi} and explained further in section 2, this ambiguity 
is of little consequence. The reason is as follows. While the 
effective action functional and, consequently, the free energy of
the free string gas are unambiguously normalized, the leading $T$
dependence of the free energy comes only from massless field theory
modes. A difference of a factor of two in the normalization is easily
absorbed in a rescaling of $\beta$ or, more precisely, the Boltzmann
constant, which has been set to unity in our use of natural units. The 
dependence of the free energy on the Planck scale massive modes is
genuinely sensitive to the factor of two because of the additional exponential
dependence on $\beta$: $F$ $\simeq$ $\beta^r e^{-\beta M^2}$, 
where $r$ is some integer characteristic of the free string gas. These 
massive modes are absent in the low energy field theory limit of the 
free string gas. Thus, while the topology of Euclidean time is indeed 
distinguished by the full string theory result the ambiguity 
is a moot point for the free energy of the massless field theory 
modes.

\vskip 0.1in
The physical consequence of imposing the ${\rm Z}_2$ twist in Euclidean time
is as follows. From Ginsparg's analysis of the moduli space of $c$$=$$1$
conformal field theories, we notice that the Hagedorn radius of the circle 
compactified bosonic string, $2 \pi \alpha^{\prime 1/2}$, coincides with 
the Kosterlitz-Thouless (K-T) point lying at the intersection of orbifold and 
circle fixed lines \cite{ginsparg}. The Kosterlitz-Thouless point is 
labelled $r_{\rm circ.}$$=$$2\pi \alpha^{\prime 1/2}$ on the fixed line of 
circle compactifications, but along the orbifold fixed line, it is 
labelled $r_{\rm orb.}$$=$$\pi \alpha^{\prime 1/2}$. This is also the 
self-dual radius of the orbifold. Thus, approaching the K-T point along the 
orbifold fixed line we find that the
K-T transition occurs at the self-dual temperature of the free string gas: 
$T_C$$=$$\pi \alpha^{\prime 1/2}$. Each of the thermodynamic potentials 
should undergo a continuous phase transition at the self-dual 
temperature exhibiting analytic behavior in the inverse temperature. 
This is precisely what we find when we evaluate the thermodynamic 
potentials: the internal energy is shown to vanish at $T_C$, and the 
Gibbs and Helmholtz free energies are minimized at the critical point.
States localized at the fixed points of the orbifold: 
$X^0$$=$$0$, $\pi r_{\rm circ}$, under 
the ${\rm Z}_2$ transformation, $X^0$$\to$$-X^0$, 
contribute a constant term to the Helmholtz free energy and entropy of the 
bosonic string gas, 
but are otherwise absent from
the internal energy, specific heat, and remaining thermodynamic potentials.
The free bosonic string gas is, of course, only of pedagogical value
since it contains a zero temperature tachyon in its mass spectrum.
This fixed-point entropy of the free bosonic string 
gas will be found to be absent in the physically meaningful case of the
heterotic string gas \cite{fermi}. 
We will find in \cite{fermi} that the general features of the duality 
transition hold also for the self-dual free heterotic string gas at 
$T_C$, but without the unphysical fixed point entropy.

\vskip 0.1in
Finally, in section 4, we derive the one-loop thermodynamic potentials 
for the free closed bosonic string gas. We should emphasize that, since
they are derived by simply taking partial derivatives with respect to
inverse temperature, or volume, of the duality invariant expression for the 
intensive string vacuum functional, the expressions for the thermodynamic 
potentials are {\em not} invariant under thermal duality transformations. 
Notice that the analyticity of the potentials as functions of inverse 
temperature follows naturally as a consequence of the analytic dependence 
of the amplitudes of perturbative string theory on any continuously-varying 
background modulus. In appendix A, we give a brief description of the 
tachyonic and massless modes in the physical state spectrum of the bosonic 
string theory, distinguishing between circle and orbifold compactifications 
of Euclidean time. We exhibit the tachyonic instability of each
momentum mode below a characteristic critical temperature, $T_n$, 
and the corresponding tachyonic instability for each winding mode beyond 
a characteristic critical temperature, $T_w$. For the closed bosonic string, 
the onset of the leading winding instability, unfortunately, occurs prior 
to removal of the leading momentum instability. As a consequence, {\em the 
thermal spectrum is tachyonic over the entire temperature range, starting at 
T=0}. Thus, the features of the thermal self-duality transition derived in 
section 4 are 
to be interpreted as strictly formal relations for the tachyonic 
bosonic string ensemble. They are a consequence of the Euclidean T-duality 
invariance of the closed bosonic string theory,
a property it shares in common with the self-dual heterotic string
theory. It is the tachyon-free free heterotic string gas which will provide 
us with a physical realization of the thermal self-duality phase 
transition, as shown in \cite{fermi}. Appendix A concludes with a 
brief summary of previous arguments in the literature which mistakenly 
confuse a tachyonic winding mode instability with a signature for a Hagedorn 
phase transition in the free string gas. We include this clarification for 
pedagogical purposes. Appendix B clarifies the precise form of the 
fixed point contribution in the entropy and Helmholtz free energy.
We review the derivation of the orbifold partition function, and the
relationship of circle and orbifold fixed lines in the c=1 moduli space
following \cite{ginsparg,polchinskibook}.

\vspace{0.3in}
\section{Aspects of Particle and String Thermodynamics}

In this paper, we will compute the generating functional of connected 
one-loop vacuum graphs in a finite temperature closed string theory.
We begin by recalling some aspects of the zero temperature 
computation of the one-loop contribution to the vacuum energy density 
\cite{poltorus}. Note that the Polyakov path integral at fixed topology, or at 
fixed order in the 
string coupling constant, is a sum over connected Riemann 
surfaces of definite topology \cite{polyakov,poltorus}. 
Note that in the functional integral formulation for perturbative
string theory, the Polyakov path integral corresponds to
$W$ $=$ $-{\rm ln} ~ Z$, 
where $Z$ $=$ $<\Omega | \Omega >$, 
and $|\Omega>$ is the interacting perturbative
closed string vacuum at zero temperature.
There is no string theory analog of 
the vacuum-to-vacuum amplitude usually computed in 
quantum field theory \cite{jackiw}. 
At zero temperature, and in the absence of background fields, the 
sum over connected Riemann surfaces of different topology gives the
string theory analog of the sum over connected 
vacuum diagrams in field theory \cite{jackiw,poltorus}:
\begin{equation}
 W[J]|_{{{\delta W}\over{\delta J}} = 0} = 
   \Gamma (0) =  {\rm sum ~ of ~ irreducible ~ connected ~ vacuum ~ 
     graphs} \quad .
\label{eq:freezero}
\end{equation}
Here $\Gamma$ is the quantum effective action functional or generating 
functional of one-particle-irreducible vacuum graphs \cite{jackiw}:
\begin{equation}
 W[J] = \Gamma (\phi_{\rm cl.}) + \int d^d x 
    {{\delta W[J]}\over{\delta J(x)}} J(x)  , \quad 
 {{\delta W[J]}\over{\delta J(x)}} = \phi_{\rm cl.} =  
    <\Omega | \phi |\Omega>_J  \quad ,
\label{eq:wgamma}
\end{equation}
where $|\Omega>$ is the interacting vacuum and $J$ an external source.
The one-loop closed string vacuum graph computed in \cite{poltorus} 
corresponds to $\Gamma(0)$: 
in the absence of background fields, $\phi_{\rm cl.} $$=$$<0|\phi|0>$$=$$0$. 
As in perturbative quantum field
theory, we calculate using a perturbation series in a small dimensionless
coupling, $g_{\rm closed}$, which, for the purposes of the
one-loop vacuum amplitude, implies that $|\Omega>$ is to be
replaced by the free string vacuum, $|0>$. 
Thus, the mass spectrum we infer from the factorization limit of the
one-loop vacuum graph is the free closed string mass spectrum. 
This has important implications for string thermodynamics since it
implies that, despite the fact that every perturbative string theory 
includes perturbative quantum gravity, the thermodynamics of the free 
string gas is expected to
have a self-consistent formulation free of gravitational instabilities, 
including the Jeans instability: free strings do not gravitate.
Loop corrections, on the other hand, are expected to be sensitive
to gravitational instabilities.

\vskip 0.1in
The effective action functional is an intensive, and dimensionless, thermodynamic 
variable. It is sometimes convenient to work with the extensive effective potential 
defined as follows \cite{jackiw,poltorus}:
\begin{equation}
\Gamma [ \phi_{\rm cl.}] = -  V_{\rm eff.} (\phi_{\rm cl.}) \cdot
\int d^{d+1} x
\quad .
\label{eq:genfn}
\end{equation}
The one-loop effective potential in the zero temperature free string
vacuum, $V_{\rm eff.}(0)$, is nothing but the one-loop contribution 
to the vacuum energy density in flat spacetime, or one-loop cosmological
constant, $\rho_0$. For bosonic string theory, a first principles 
computation of $\rho_0$ from the Polyakov path integral was given 
in \cite{poltorus}, where the unambiguous normalization of the vacuum
energy density in string theory was first noted. The one-loop 
string path integral is pure gauge as a consequence of two dimensional 
general coordinate invariance \cite{polyakov}. Its normalizability in
a spacetime infrared finite string theory follows from the existence of a 
gauge-invariant world-sheet regulator preserving the Weyl invariance 
of the measure in the string path integral \cite{poltorus,rg}.

\vskip 0.1in
Note that the formal infrared divergences of the bosonic string path 
integral as a consequence of the tachyon are being ignored here,
in the same spirit as the discussion in the introduction. The reader 
should interpret these statements in the context of their implications 
for a tachyon-free, 
and infrared finite, ground state of the self-dual heterotic string. The 
world-sheet technicalities are otherwise similar. From a spacetime 
perspective, a tachyon-free heterotic string theory is both ultraviolet 
and infrared finite. The vacuum energy density will be both finite and 
calculable as a function of spacetime moduli, the continuously varying 
background parameters of a consistent string ground state \cite{gv}. In 
a spacetime supersymmetric ground state, the vacuum energy 
density will of course vanish, but the contributions from spacetime
bosonic, and spacetime fermionic, modes are {\em separately} finite and calculable. 
A word about dimensions is in order. Since $Z$ 
and $W$ are both dimensionless, the effective potential or vacuum energy
density has dimensions of $L^{-(d+1)}$ in $d$ spatial dimensions. We use 
the natural units $\hbar$$=$$c$$=$$k_B$$=$$1$, where $k_B$ is Boltzmann's
constant. 

\vskip 0.1in
Let us now recall some of the standard wisdom in quantum statistical
mechanics and finite temperature field theory \cite{dolan,daskap}.
Consider the time independent quantum statistical mechanics
of an equilibrium ensemble of particles with Hamiltonian, $H$, at
fixed temperature $T$, and confined to a fixed spatial volume $V$.
The canonical partition function, $Z$$=$${\rm  Tr} e^{-\beta {\hat{H}}}$,
the Helmholtz free energy, $F$, internal
energy, $U$, pressure, $P$, entropy, $S$, and specific heat at 
constant volume, $C_V$, of the canonical ensemble are 
defined by the usual thermodynamic identities:
\begin{equation}
F = - T {\rm ln} Z , \quad 
U =  T^2 {{\partial}\over{\partial T}} {\rm ln} Z ,  \quad
P =  - \left ( {{\partial F }\over{\partial V}} \right )_T  ,
\quad
S =  - \left ( {{\partial F }\over{\partial T}} \right )_V  ,
\quad
C_V =  T \left ( {{\partial S}\over{\partial T}} \right )_V \quad .
\label{eq:can}
\end{equation}
All thermodynamic functions in quantum statistical mechanics 
can be derived, in principle, from a knowledge of density matrix 
elements of the 
evolution operator, $e^{-\beta {\hat{H}}}$. In finite temperature 
quantum field theory, the ill-defined nature of the trace requires that
we replace these definitions with normalized expectation values of 
observables in a thermal ensemble. In either case, the evolution 
equation satisfied by the density matrix suggests that we 
introduce a fictitious evolution parameter, $\tau$, with range, 
$0$ $<$ $\tau$ $\le$ $\beta$. It is natural 
to associate $\tau$ with an auxiliary imaginary time, $\tau$$=$$-it$.
Thus, we can interpret the equilibrium Greens functions of an
interacting $d$-dimensional thermal field theory as Greens functions 
in an interacting $d$$+$$1$-dimensional Euclidean field theory, with 
Euclidean time an interval of size $\beta$. Notice that \lq\lq time reversal 
invariance" is a symmetry with no natural physical analog in the 
Euclidean time prescription. We will return to this point in the conclusions.

\vskip 0.1in
Operators in the Euclidean field theory are defined via the Heisenberg
representation, $\phi(\tau,x)$ $=$ $e^{\tau H_0} \phi (x) e^{-\tau H_0}$,
with ensemble averages defined in the non-interacting Fock space.
The fundamental object in the thermal field theory is the propagator which
must satisfy a periodicity, or aperiodicity, condition in Euclidean time
as a consequence of cyclicity of the trace \cite{daskap}:
\begin{equation}
G_{\beta}(\tau,\tau') = Z^{-1}(\beta) \left 
  [ {\rm Tr} ~ e^{-\beta H} ~ {\rm P}_\tau
( \phi (x,\tau) \phi^{\dagger} (x',\tau^{\prime})) \right ] , \quad
G_{\beta}(0,\tau) = \pm G_{\beta}(\beta,\tau) \quad .
\label{eq:per}
\end{equation}
The choice of sign corresponds, respectively, to fields with
integer, or half-integer, spin.
The thermal Greens function for a free scalar field in $d$
dimensions can be Fourier expanded in the basis:
\begin{equation}
G_{\beta}(\tau,\tau') =
  {{1}\over{\beta}} \sum_{n} \int {{d^d k}\over{(2\pi )^d}}
   e^{-i(\omega_n \tau + {\bf k}\cdot{\bf x}) }
     G_{\beta} (\omega_n, {\bf k}), \quad  \omega_n=2n\pi/\beta , 
  ~ n \in {\rm Z}.
\label{eq:green}
\end{equation}
Conversely, the free fermion is expanded in a Fourier basis with 
odd frequencies: $\omega_n$$=$$(2n+1)\pi/\beta$, ensuring 
the aperiodicity of the Greens function in Euclidean time. 
The Euclidean time prescription can be equivalently stated in 
terms of the Euclidean functional integral, where we restrict the 
path integral to Fourier modes that are, respectively, 
periodic or aperiodic in Euclidean time, for fields having,
respectively, bosonic or fermi spacetime statistics \cite{daskap}:
\begin{equation}
Z(\beta) =  {\rm Tr} ~ e^{-\beta H} = \int d \Phi < \Phi | e^{-\beta H}
| \Phi > ~ \equiv ~ N \int_{\rm periodic}
[ {\cal D} \phi ] e^{- \int_0^{\beta} d \tau \int d^d x
{\cal L} (\phi({\bf x}, \tau) ) } \quad .
\label{eq:parti}
\end{equation}
Thus, the equilibrium behavior of a finite temperature field theory 
in $d$ spatial dimensions has a field theoretic formulation in the 
Euclidean spacetime, 
$R^d $$\times$(a one-dimensional compact target space of
volume $\beta$), where $\beta$ is the inverse temperature.  

\vskip 0.1in
There is an ambiguity in the Euclidean time prescription that becomes 
apparent when we try to apply it to finite temperature string theory. 
Surprisingly, this has not been pointed out prior to the appearance of
our work \cite{us}.
The topology of a compact one-dimensional target space is either that
of a circle, $S^1$, with inverse temperature $\beta$ identified as the 
circumference, $2\pi r_{\rm circ}$, or, that of an orbifold, $S^1/Z_2$, 
where the $Z_2$ acts on the circle as a reflection: $X^0$$\to$$-X^0$,
with fixed points at $X^0$$=$$0$ and $X^0$$=$$\pi r_{\rm circ}$,
and $\beta$ identified as the interval length, $\pi r_{\rm circ}$. Being 
extended objects, free strings can explore the global topology of the 
space in which they live. Consequently, both the free string mass 
spectrum, and the free energy of the free string gas, are expected to 
distinguish between the two possible topologies for Euclidean time.

\vskip 0.1in
Further, recall that the prescription for finite temperature quantum 
field theory \cite{dolan} requires that we sum over the same set of connected 
vacuum graphs as at zero 
temperature, but taking care to replace all integrals in frequency space 
with infinite sums over quantized Matsubara frequencies \cite{daskap}. 
Either circle or orbifold topologies for Euclidean time result in an infinite 
sum over quantized thermal frequencies, since this is simply a property of 
any compact target space.
Moreover, all of the modes in the bosonic string spectrum are spacetime bosons,
and consequently the string functional integral is already restricted to modes 
that are periodic in Euclidean time, quite independent of topology. 
The ambiguity in the Euclidean time prescription remains. But as already
mentioned in the Introduction, the choice of orbifold toplogy has the
satisfying physical consequence that the thermal duality transition 
can be concretely identified with the Kosterlitz-Thouless continuous
phase transition. In the physically meaningful case of the heterotic
string gas, the ${\rm Z}_2$ action is essential in eliminating the 
tachyon while preserving modular invariance and the correct zero 
temperature limit \cite{fermi}. 

\vskip 0.1in
Our starting point for a discussion of closed string thermodynamics in
this paper is the world-sheet path integral representation of the
generating functional 
of connected one-loop vacuum string graphs, $W$$=$$W(\beta)$, in the
embedding space $R^d$$\times$$S^1/Z_2$, and with interval length identified as
inverse temperature, $\beta$ \cite{polyakov,poltorus}:
\begin{equation}
W(\beta) \equiv {\rm ln}~ Z , \quad
F(\beta) = -W/\beta , \quad  
\rho(\beta) = -W/V \beta , \quad  
      U(\beta) = - T^2 \left ( {{\partial W}\over{\partial T}} \right )_V =
             \left ( {{\partial W}\over{\partial \beta}} \right )_V 
\quad .
\label{eq:freene}
\end{equation}
$F$ is the Helmholtz free energy of the ensemble of free strings, 
$U$ is the internal energy, and $\rho$ is the finite temperature 
effective potential. Notice that $\rho$ is the finite temperature analog of 
the one-loop vacuum energy density or cosmological constant, $\rho_0$.
The inverse temperature, $\beta$, plays the role of a spacetime
modulus, a continuously-varying background parameter. Thus, finite temperature 
string theory describes a one-parameter family of consistent ground states 
of string theory, characterized by the $\beta$ dependent effective potential.
We emphasize that the thermodynamic potentials of the 
free string ensemble are obtained directly from the generating 
functional of connected vacuum graphs, ${\rm ln}$ ${\rm Z}(\beta)$, as
opposed to ${\rm Z}(\beta)$. Thus, we never need address the troubling 
issue of defining the thermodynamic limit of the canonical ensemble, since 
we never compute the canonical partition function for string states directly 
in the path integral framework. 

\vskip 0.1in
We should warn the
reader that it is conventional in the string theory literature--- and we will
stick with this convention, to refer to the modular invariant integrand 
of the string vacuum functional, or free energy, as the 
\lq\lq partition function", usually denoted by ${\rm Z}(\tau,{\bar{\tau}})$. 
The reason is that the $q$${\bar{q}}$ $=$ $e^{-4\pi\tau_2}$ expansion of 
${\rm Z}(\tau, {\bar{\tau}})$ is correctly interpreted as the level 
expansion in string theory: the $n$th power of $q$${\bar{q}}$ in the 
Taylor series expansion gives the degeneracy of the $n$th mass level in the string mass 
spectrum, upon imposition of the level matching condition on physical states:
$L_0$$=$${\bar{L}}_0$. 
Terms in the Taylor expansion with unequal powers of $q$ and ${\bar{q}}$ do not 
meet the level matching condition, and correspond to unphysical states
in the mass spectrum. They drop out during the process of integrating
over $\tau_1$, and do not contribute to the free energy of the free string gas. 
A small clarification is in order. There is a surprisingly extensive literature 
on unphysical tachyons and their role in determining the asymptotics of the function 
${\rm Z}(\tau, {\bar{\tau}})$ as, for example, in ref.\ \cite{kani}. We should 
clarify that, in our work, when we refer to a tachyonic instability we mean a 
physical tachyonic 
mode which meets the level matching conditions. Unphysical tachyons do enter the
BRST algebra of a covariant string theory, but we have not found any useful 
role for these states, even in the case of the thermal fermionic strings 
discussed in \cite{fermi}. 

\section{Effective Potential and Free Energy}

Consider an equilibrium
ensemble of closed bosonic strings at fixed temperature in $d$$=$$25$
spatial dimensions. The strings can be assumed to be confined to a
box-regularized spatial volume:
\begin{equation}
 V = L^{25} (2\pi \alpha^{\prime})^{25/2} \quad .
\label{eq:vol}
\end{equation}
Since the vacuum string functional of interest, $W(\beta)$ $=$ ${\rm ln}$ $Z(\beta)$, 
is an intensive thermodynamic variable, we will find that the precise
regularization of the spatial volume is, fortunately, of no consequence.
We have defined the inverse temperature to have
dimensions of length, or inverse energy, working with the natural units
$\hbar$ $=$ $c$ $=$ $1$, and with Boltzmann's constant set to unity.
The normalized generating functional for 
connected one-loop vacuum string graphs in the Euclidean embedding space 
$R^{25}$$\times$$S^1/Z_2$ can be computed as shown in 
\cite{polyakov,poltorus,polchinskibook,us}. This directly gives the 
Helmholtz free energy of the free closed bosonic string gas: 
$F$$=$$-W/V\beta$, the basic thermodynamic potential from which we 
can derive all of string thermodynamics in the canonical formalism. Our 
results can be compared with the partial insights gained in the
previous works 
\cite{huangw,carlitz,bowick,poltorus,mclain,longs,tan,sathia,aw}.
As emphasized in the introduction, the important differences in 
our approach come from taking seriously the identification of 
the Polyakov path integral as the world-sheet representation of 
${\rm ln}$ ${\rm Z}(\beta)$ and requiring, in addition,   
that the generating functional of connected string graphs
transform correctly under Euclidean T-duality transformations. For the 
closed bosonic string 
gas, this implies a precise invariance of $W(\beta)$ under thermal 
duality transformations: $\beta$$\to$$\beta_C^2/\beta$.

\subsection{Normalization and Phase of the Effective Potential}

We begin with the world-sheet representation of the generating
functional of connected one-loop vacuum string graphs in closed
bosonic string theory at finite temperature. Since the derivation
of the one-loop string path integral for closed bosonic string 
theory compactified on the spatial orbifold $S^1/Z_2$ can be 
found in the references \cite{polyakov,poltorus,polchinskibook},
we will simply write down the result \cite{us}:
\begin{eqnarray}
W_{\rm bos.} (\beta) =&&
 \half \int_{\cal F}
{{|d\tau|^2}\over{4\pi\tau_2^2}} \left ( (2\pi\tau_2)^{-23/2}
   |\eta(\tau)|^{-46} \right ) \cr 
&&\times \left [ {{1}\over{\eta{\bar{\eta}}}}
\sum_{n,w \in Z} q^{ {{1}\over{2}}{{\alpha' p_L^2}\over{2}} }
    {\bar q}^{{{1}\over{2}} {{\alpha' p_R^2}\over{2}}} 
+ \left ( | \Theta_3 \Theta_4 | + | \Theta_2 \Theta_3 | + | \Theta_2 \Theta_4 |
\right ) \right ] .
\label{eq:boso}
\end{eqnarray}
$W(\beta)$ is an intensive thermodynamic variable.
Dividing out by the volume of Euclidean spacetime gives the 
dimensionful finite temperature effective potential:  
$\rho(\beta)$ $=$ $- W/V\beta$. The various factors in $\rho(\beta)$ may 
be understood as follows. The factor of 
$\beta$ $=$ $\pi r_{\rm circ.}$ arises 
from the integration over Euclidean time, the interval length of the orbifold
$S^1/{\rm Z}_2$. The box-regularized spatial volume arises from integration 
over the zero modes of the $25$ spatial embedding coordinates. 
The factor of ${{1}\over{2}}$ is the $Z_2$ symmetry factor 
obtained when we project to the subspace of string modes invariant
under the reflection, $X^0$$\to$$-X^0$. The relative signs of the
different terms in the square brackets are 
determined by modular invariance.
Here, $ p_{L(R)}$$={{ \pi n}\over{\beta}} 
\pm {{w \beta }\over{ \pi \alpha' }}$.
We can introduce a dimensionless inverse temperature 
(radius) defining $x$ $\equiv$ $r(2/\alpha^{\prime})^{1/2}$, 
with $\beta$ $=$ $\pi (\alpha^{\prime}/2)^{1/2} x$. The 
dimensionless quantized momenta live in
a $(1,1)$ dimensional Lorentzian self-dual lattice 
\cite{polchinskibook}:
\begin{equation}
\Lambda^{(1,1)} :  \quad\quad ({{\alpha^{\prime}}\over{2}})^{1/2}
(p_L , p_R) \equiv   (l_L , l_R) = ( 
{{n}\over{x}} + {{wx}\over{2}}   , 
{{n}\over{x}} - {{wx}\over{2}}  )  \quad , 
\label{eq:dimless}
\end{equation} 
The momentum summation is manifestly invariant under
thermal duality,
$\beta$$\to$$\pi^2 \alpha^{\prime}/\beta$, or $x$ $\to$ $2/x$,
simultaneously interchanging the dummy indices, 
$n $ $\to$ $w$.

\vskip0.1in
We begin with a simple analysis of the massless 
states in the physical state 
spectrum of the orbifold, clarifying the necessity for
a ${\em Z}_2$ twist in the Euclidean time prescription for the
thermal bosonic string.
Consider the mass formula for physical states in the untwisted
sector of the orbifold:
\begin{equation}
({\rm mass})^2 =  {{2}\over{\alpha^{\prime } }}
    ( -2 + \half {\bf l}_L^2 + 
\half {\bf l}_R^2 + N_0 + {\bar N}_0  + 
 N_{\mu} + {\bar N}_{\mu} ) \quad ,
\label{eq:massesm}
\end{equation}
where we are required to project onto the subspace of states that are
invariant under the reflection: state with $N_0$ $+$ ${\bar{N}}_0$ equal to
an even integer, and lattice vectors invariant under the reflection,
$(n,w)$ $\leftrightarrow$ $(-n,-w)$.
The level-matching condition for physical states takes the form: 
\begin{equation}
2nw + N_0 - {\bar{N}}_0 + N_{\mu}-{\bar{N}}_{\mu}=0 , \quad
\mu=1, ~ \cdots, ~ 25. 
\label{eq:lmm}
\end{equation}
The zero point energy in the twisted sectors differs by $+{{1}\over{16}}$,
and the oscillator moding will be half-integral. There are no massless
states in the twisted sectors of the orbifold.

\vskip 0.1in
Denote a generic state in the bosonic string spectrum by 
$|\alpha_{-1} \cdots {\bar \alpha}_{-1} \cdots
|k , {\bar k};(n,w)>$.
Notice that the $Z_2$ twist in Euclidean time 
breaks the $U(1)$$\times$$U(1)$ symmetry
present at generic radius in the circle compactification, 
thereby removing the Kaluza-Klein vector states:
$\alpha^{\mu}_{-1} {\bar \alpha}^0_{-1} |k^{\mu},0;(0,0)>$,
and $\alpha^{0}_{-1} {\bar \alpha}^{\mu}_{-1} |0,{\bar k}^{\mu};(0,0)>$, 
$\mu$$=$$1$, $\cdots$, $25$, from the massless spectrum.
A scalar massless modulus, $\alpha^{0}_{-1} 
{\bar \alpha}^{0}_{-1} |0,0;(0,0)>$,
remains, a marginal perturbation of the world-sheet 
conformal field theory corresponding to a change of radius 
(inverse temperature). 
The remaining massless states in the physical state
spectrum are the $25$-dimensional graviton, antisymmetric tensor field, 
and dilaton, of the bosonic closed string theory. 

\vskip 0.1in
It is instructive to compare our expression in Eq.\ (\ref{eq:boso})
for the Helmholtz free energy with Eq.\ (33) of \cite{poltorus}, where the
high temperature limit of the
generating functional of connected one-loop vacuum string graphs 
was computed in the path integral formulation. 
The normalized, and modular invariant, measure 
for moduli in Eq.\ (\ref{eq:boso})
is exactly as derived in 
\cite{poltorus,polchinskibook}. It corresponds to the world-sheet metric,
$ds^2$ $=$ $|d(\sigma^1 + \tau \sigma^2)|^2$, where $\tau = \tau_1 + i \tau_2$
is the complex modulus for the torus. The integration
region, ${\cal F}$, is the fundamental domain of the torus \cite{polchinskibook}.
The factor in round brackets is the modular invariant contribution from free 
string periodic oscillators in $23$ non-compact transverse directions.
Here, $q$$=$$e^{2\pi i \tau}$. As mentioned above, the integration 
over $\tau_1$ imposes the level matching condition, $L_0$ $=$ ${\bar L}_0$, 
projecting out all states other than the physical states in the
mass spectrum of the free closed bosonic string.

\vskip 0.1in
The thermal duality invariant contribution to ${\rm Z}(\tau,{\bar \tau})$ 
from winding and momentum modes is easily derived in the operator formalism 
using current algebra methods \cite{ginsparg,klt}. The integers $n$ and $w$ 
label, respectively, momentum and winding
modes of the string in Euclidean time. For closed bosonic strings in
an embedding space with
one compact dimension, the physical state condition reads: 
$2nw$$+$$N_0$$-$${\bar{N}}_0$$+$$N_{\mu}$$-$${\bar{N}}_{\mu}$$=$$0$, 
$\mu$$=$$1$, $\cdots$, $25$. The ${\rm Z}_2$ orbifold twist implies 
that in the untwisted sector we project to the subspace of states 
for which $N_0$ $+$ ${\bar N}_0$ is even, keeping only the
symmetric linear combinations of the $(+n,0)$ and $(-n,0)$
momentum states, and the $(0,+w)$ and $(0,-w)$ winding states.
The zero point energy in the twisted sectors is higher than that in the
untwisted sector by the factor $+{{1}\over{16}}$. There are no massless
modes in the twisted sectors. We note that the tachyonic states in the twisted sector
lead to subleading divergences due to the higher zero point energy. 
We reiterate that the fixed point contributions are absent in
the internal energy and successive partial derivatives with respect to
temperature. They do appear in the Helmholtz free energy and 
entropy of the bosonic string
gas, but will be absent in the corresponding expressions for the heterotic
string gas.

\vskip 0.1in
A Poisson resummation of the factor in square brackets in Eq.\ 
(\ref{eq:boso}) puts it in a manifestly modular invariant form suitable 
for comparison with the path integral derivation in \cite{poltorus}:
\begin{equation}
\left [ \cdots \right ] =
 {{1}\over{ \eta{\bar{\eta}} }}
\left [ (2\tau_2)^{-1/2}
\sum_{n,w \in Z} e^{ - {{\pi x^2 }\over{2 \tau_2 }}|n - w \tau |^2  }
          e^{\pi i n w x^2 }
+ \left ( | \Theta_3 \Theta_4 | + | \Theta_2 \Theta_3 | + | \Theta_2 \Theta_4 |
\right ) \right ]
\quad .
\label{eq:poiss}
\end{equation}
The path integral derivation of the contribution from winding and 
momentum modes is as follows. In 
the high temperature limit, we can restrict
the $p^0$ summation to modes winding in the $\sigma^2$ direction. 
We decompose $X^{0}$$=$$y^0$$+$$X^0_{\rm cl.}$, where $X^0_{\rm cl.}$ 
is a classical solution to the equations of motion periodic in the 
$\sigma^2$ direction \cite{poltorus}:
\begin{equation}
X^{\mu} (\sigma^{1} , \sigma^2 + 2\pi ) =
y^{\mu} (\sigma^{1} , \sigma^2) + 2 n \beta \sigma^2 \delta^{\mu}_0
, \quad S(X,g_{ab}) = S(y, g_{ab}) +
  {{ \pi n^2 x^2 }\over{2 \tau_2}} , \quad n \in {\rm Z}
\quad .
\label{eq:wind}
\end{equation}
$S$ is the classical action for the Polyakov string.
We have used the fact that in the world-sheet metric defined earlier,
${\sqrt g} g^{22}$ $=$ $1/\tau_2$. 
The extra term in the world-sheet action contributes the
exponential: $\sum_{n=-\infty}^{\infty} e^{-n^2 \beta^2
/\pi \alpha^{\prime} \tau_2 }$, to the expression for the
one-loop vacuum energy density 
\cite{poltorus}. Generalizing to modes which wind in
both $\sigma^1$ and $\sigma^2$ directions, and referring 
to the full world-sheet metric, we obtain:
\begin{eqnarray}
X^{\mu} (\sigma^{1} + 2 \pi, \sigma^2 + 2\pi ) &&=
y^{\mu} (\sigma^{1} , \sigma^2) + 2 n \beta \sigma^2 \delta^{\mu}_0
+ 2 w \beta \sigma^1 \delta^{\mu}_0
\cr
S(X,g_{ab}) &&= S(y, g_{ab}) +
{{\pi x^2 }\over{2 \tau_2 }} \left (
n^2 + |\tau|^2 w^2 -2 \tau_1 nw \right )^2  
\quad ,
\label{eq:windd}
\end{eqnarray}
with $w$, $n$ $\in$ ${\rm Z}$.
Performing the Gaussian path integrals as before gives the terms in
the exponential, in agreement with both Eq.\ (\ref{eq:poiss}) and 
Eq.\ (\ref{eq:boso}). 
Notice that modular invariance is manifest in Eq.\ (\ref{eq:poiss}) 
while thermal duality invariance becomes manifest in the form  
given in Eq.\ (\ref{eq:boso}).
Thermal duality 
interchanges the low and high temperature phases of string theory, 
and like modular invariance, it requires the presence of both species 
of modes in the vacuum energy functional \cite{polchinskibook}. 

\vskip 0.1in
To summarize, the thermal duality invariant result for the 
generating functional of connected one-loop vacuum graphs 
in the embedding space $R^{25}$$\times$$S^1/Z_2$
given in Eq.\ (\ref{eq:boso}) can be written in path integral form 
as follows:
\begin{equation}
W (\beta) = (V \beta)^{-1} \cdot \sum_{{\bf p}_L, {\bf p}_R} 
(-1)^{ \half ({\bf l_L^2 }-  {\bf l_R^2} )x^2} 
\int {{[dX] [dg_{ab}]}\over{ {\rm Vol(gauge)} }} 
e^{-S (X,g) - {{\nu_0}\over{4\pi}} \int d^2 \sigma R_g {\sqrt{g}} 
   - \mu_0 \int d^2 \sigma {\sqrt{g}} } 
\quad .
\label{eq:pathi}
\end{equation}
Here, $\mu_0$ is the bare world-sheet cosmological constant
which renormalizes to zero in the Weyl anomaly free
critical string \cite{polyakov,poltorus}. $R_g$ is the intrinsic
curvature of the world-sheet, and the term proportional to the Euler
character will, of course, vanish for surfaces with the topology of a torus.
We have written $W$ in a form which clarifies that multi-loop contributions 
to the vacuum energy density are also required to be
thermal duality invariant:
we sum over momentum sectors weighted by arbitrary temperature dependent
phases, subject to the preservation of thermal duality invariance. 
Within each sector, we perform the path integral in a manifestly 
reparameterization invariant way preserving Weyl invariance
\cite{poltorus}. This is an intriguing observation since, in 
practice, it may be easier to impose thermal duality on the phase of 
the multiloop path integral as opposed to modular invariance.  

\subsection{High Temperature Effective Potential and Holography}

The generating functional for connected one-loop vacuum string 
graphs is invariant under a thermal duality transformation: 
$W(T)$ $=$ $W(T^2_c/T)$,
with self-dual temperature, $T_c$ $=$ $1/\pi \alpha^{\prime 1/2}$.
As pointed out by Polchinski \cite{polchinskibook}, we can infer 
the following thermal duality relation which holds for both the 
Helmholtz free energy, 
$F(T)$ $=$ $-T \cdot W(T)$, and the effective potential, 
$\rho(T)$ $=$ $-T \cdot W(T)/V$:
\begin{equation}
F(T)  = {{T^2}\over{T_C^2}} F({{T_C^2}\over{T}}) ,
\quad \quad \rho(T) = {{T^2}\over{T^2_C }} \rho({{T_C^2}\over{T}})
\quad .
\label{eq:thermi}
\end{equation}
Consider the high temperature limit of this expression:
\begin{equation}
\lim_{T \to \infty } \rho(T) =
\lim_{T \to \infty}
{{T^2}\over{T^2_C }} \rho({{T_C^2}\over{T}})
 =  \lim_{(T_C^2/T^2) \to 0} {{T^2}\over{T^2_C }} 
\rho({{T_C^2}\over{T}})
 =  {{T^2}\over{T^2_C }} \rho_0 
\quad ,
\label{eq:thermasy}
\end{equation}
where $\rho_0$ is the cosmological constant, or vacuum energy 
density, at zero temperature. Likewise, at high temperatures, 
the free energy grows as the square of the temperature. Thus, 
growth in the number of degrees of freedom at high temperature
in the gas of free closed bosonic strings is only as fast as in 
a {\em two-dimensional} field theory. This is significantly 
slower than the $T^{26}$ growth of the high temperature degrees 
of freedom in the low energy finite temperature field theory.
We comment that the coincidence of the self-dual point of the
orbifold fixed line with the Hagedorn point of the bosonic string
implies that the holographic relation holds at the Hagedorn point, 
as originally conjectured in \cite{aw}. 

\vskip 0.1in
Notice that the prefactor in the high temperature relation is 
unambiguous, a consequence of the normalizability of the generating
functional of one-loop vacuum string graphs in string theory 
\cite{poltorus}. The pre-factor, $\rho_0/T_C^2$, is also background 
dependent: it can be computed as a continuously varying function of 
the background fields upon compactification to lower spacetime 
dimension \cite{gv}. The relation in Eq.\ (\ref{eq:thermasy}) is 
unambiguous evidence of the holographic nature of perturbative string 
theory: a reduction in the degrees of freedom of a self-dual closed 
string theory at high temperatures, or short distances \cite{aw}.
An identical holographic relation holds for a physical realization 
of the self-dual gas of free closed strings: the tachyon-free high 
temperature gas of free heterotic strings with gauge group 
$SO(16)$$\times$$SO(16)$ \cite{us,fermi}.

\section{Kosterlitz-Thouless Self-Duality Phase Transition}

In this section we show how to systematically derive the full set of 
thermodynamic potentials for the canonical ensemble of free strings.
We will verify from the analytic behavior of the thermodynamic 
potentials as functions of inverse temperature that the phase transition 
in a self-dual gas of free closed strings is of the Kosterlitz-Thouless 
type: the infinite hierarchy of thermodynamic potentials exhibits 
analytic behavior as a function of temperature at the critical point 
\cite{kosterlitz}. 
As a consequence of thermal duality, we will find that
the internal energy vanishes precisely at the 
self-dual temperature. The internal energy is a monotonically 
increasing function of temperature, crossing from negative values
at low temperature to positive values at temperatures above $T_C$.
The low temperature phase is dominated by the Kaluza-Klein states.
In the high temperature phase, long winding strings become
energetically feasible, and are entropically favoured. 
We remind the reader that the thermal spectrum of the bosonic
string is tachyonic at all temperatures starting from zero.
In the tachyon-free physical realization of a self-dual free 
closed string gas, namely the finite temperature heterotic 
ensemble described in \cite{fermi}, we expect to find that 
both the Helmholtz and Gibbs free energies are minimized at $T_C$, 
while the specific heat is {\em positive-definite}. These features
could be taken as indicators of thermodynamic stability for the 
self-dual gas of free heterotic strings.

\vskip 0.1in
We begin with the modular and thermal duality invariant expression for
the generating functional of finite temperature one-loop vacuum graphs 
in the closed bosonic string:
\begin{equation}
W (\beta) = \half \int_{\cal F}
{{|d\tau|^2}\over{4\pi\tau_2^2}} (2\pi\tau_2)^{-23/2}
   |\eta(\tau)|^{-48} 
\left [ \sum_{n,w} q^{ {{1}\over{2}}{{\alpha' p_L^2}\over{2}} }
    {\bar q}^{{{1}\over{2}} {{\alpha' p_R^2}\over{2}}} +
\left ( | \Theta_3 \Theta_4 | + | \Theta_2 \Theta_3 | + | \Theta_2 \Theta_4 |
\right ) \right ] .
\label{eq:free}
\end{equation}
Recall the relation, 
$\beta$ $=$ $\pi x ({{\alpha^{\prime}}\over{2}})^{1/2}$.
Referring to the thermodynamic definitions given in Eq.\ (\ref{eq:can}), 
it is evident that the expressions we will derive for the thermodynamic 
potentials are not invariant under thermal duality transformations.
The expression for the internal energy of the gas of free closed
bosonic strings takes the form:
\begin{equation}
 U(\beta) = - \left ( {{\partial W}\over{\partial \beta }} \right )_V =
 \half \int_{\cal F}
{{|d\tau|^2}\over{4\pi\tau_2^2}} (2\pi\tau_2)^{-23/2}
   |\eta(\tau)|^{-48}
{{4\pi \tau_2}\over{\beta}}
\sum_{n,w} 
  \left ( {{w^2 x^2}\over{4}} - {{n^2}\over{x^2}} \right )
\cdot 
q^{ {{1}\over{2}}{\bf l}_L^2 }
     {\bar q}^{{{1}\over{2}}{\bf l}_R^2 }
\quad .
\nonumber\\
\label{eq:term}
\end{equation}
$U(\beta)$ vanishes precisely at the self-dual temperature,
$T_c$$=$$1/\pi\alpha^{\prime 1/2}$, $x_c$$=$${\sqrt{2}}$, where the 
internal energy contributed by winding sectors cancels that
contributed by momentum sectors. The internal energy changes sign at
$T$ $=$ $T_C$, transitioning from negative values at low temperature 
to {\em positive} values at high temperature. 
The balance between energy and entropy shifts at the self-dual
temperature, as seen from the definition of the Helmholtz function: 
$F$ $=$ $U$ $-$ $TS$.
The positive contribution to the internal energy from a winding 
sector implies that while the creation of a long winding string is 
costly in energy, it becomes entropically favoured at 
high temperature.

\vskip 0.1in
A clarification about the zero temperature limit is in order.
From Eq.\ (\ref{eq:free}), the contributions to the free energy 
from winding sectors are 
exponentially damped for small $\beta$. Conversely, the 
contributions from momentum sectors are exponentially damped 
at large $\beta$. 
At the self-dual temperature, winding and momentum sectors
contribute equally to $F$.
Notice that the asymptotic value of the expression for the
free energy in the $\beta$$\to$$\infty$ limit includes the 
fixed point contributions in Eq.\ (\ref{eq:free}). In order
that the asymptotic value match correctly
with the expected zero temperature result: 
$F_0$ $=$ $V\rho_0$, the noncompact limit has to be defined 
with care, suitably resolving the orbifold singularity.

\vskip 0.1in
It is easy to demonstrate the analyticity of infinitely many 
thermodynamic potentials in the vicinity of the critical point.
It is convenient to define:
\begin{equation}
[d \tau ] \equiv \half 
\left [ {{|d\tau|^2}\over{4\pi\tau_2^2}} (2\pi\tau_2)^{-23/2}
   |\eta(\tau)|^{-48} e^{2\pi i nw \tau_1} \right ] ,
\quad y(\tau_2;x) 
   \equiv 2\pi \tau_2 \left ( {{n^2}\over{x^2}} + {{w^2 x^2}\over{4}}
  \right ) \quad . 
\label{eq:vars}
\end{equation}
Denoting the $m$th partial derivative with 
respect to $\beta$ at fixed volume by
$W_{(m)}$, $y_{(m)}$, 
and setting $x$ $\equiv $ $\alpha \beta$,
we note that the higher derivatives of
the generating functional
take the simple form: 
\begin{eqnarray}
W_{(1)} =&& \sum_{n,w} \int_{\cal F} [d\tau] e^{-y} (-y_{(1)}) 
\cr
W_{(2)} =&&
 \sum_{n,w} \int_{\cal F} [d\tau] e^{-y} 
(-y_{(2)} + (-y_{(1)})^2 ) 
\cr
W_{(3)} =&& \sum_{n,w} \int_{\cal F} [d\tau] e^{-y} 
(-y_{(3)} - y_{(1)} y_{(2)} +  (-y_{(1)})^3 ) 
\cr
 \cdots =&& \cdots 
\cr
W_{(m)} =&& \sum_{n,w} \int_{\cal F} [d\tau] e^{-y} 
(-y_{(m)} - \cdots +  (-y_{(1)})^m ) 
\quad .
\label{eq:effm}
\end{eqnarray}
Referring back to the definition of $y$, it is easy to see that the
generating functional and, consequently, the full set of thermodynamic
potentials is analytic in $x$. Notice that third and higher derivatives 
of $y$ are determined by the momentum modes alone:
\begin{equation}
y_{(m)} = (-1)^m n^2 {{(m+1)! }\over{x^{m+2}}} , \quad m \ge 3 \quad . 
\label{eq:ders}
\end{equation}
For completeness, we give explicit results for the first few 
thermodynamic potentials:
\begin{equation}
F = - {{1}\over{\beta}} W_{(0)} , \quad 
U = - W_{(1)} , \quad 
S = W_{(0)} - \beta W_{(1)} , \quad 
C_V = \beta^2 W_{(2)} , \cdots \quad . 
\label{eq:thermodl}
\end{equation}
Let $2\pi \tau_2$ $\equiv$ $t$. 
The entropy is given by the expression:
\begin{equation}
S = \sum_{n,w} \int_{\cal F} [d\tau] e^{-y} 
  \left [ 1 + 2 t ( - {{n^2}\over{x^2}} + {{w^2 x^2}\over{4}} ) 
\right ] +  S_0 \quad ,
\label{eq:entropy1}
\end{equation}
where $S_0$ denotes the fixed point contributions to the 
entropy. For the specific heat at constant volume, we have:
\begin{equation}
C_V =  \sum_{n,w} \int_{\cal F} [d\tau] e^{-y} 
  \left [   4 t^2 ( - {{n^2}\over{x^2}} + {{w^2 x^2}\over{4}} )^2 
   - 2t ( 3 {{n^2}\over{x^2}} + {{ w^2 x^2}\over{4}} ) 
\right ] .
\label{eq:spc}
\end{equation}
It is easy to check the sign of the specific heat at criticality
for an infrared finite self-dual ensemble of closed strings. Since
the first term in square brackets will not contribute at $T_C$,
a stable thermodynamic ensemble with positive specific heat
requires {\em positive} Helmholtz free energy.
This is precisely as found for the finite temperature heterotic 
string with gauge group $SO(16)$$\times$$SO(16)$ \cite{fermi}. 
Due to an excess of spacetime fermions over spacetime bosons in the
physical state spectrum, it has positive free energy.
 
\vskip 0.1in
Finally, consider the fixed point contributions in the expressions 
for both the free energy, and in the entropy. The Helmholtz 
free energy takes the form:
\begin{equation}
F (\beta) = - \half {{1}\over{\beta}} \int_{\cal F}
{{|d\tau|^2}\over{4\pi\tau_2^2}} (2\pi\tau_2)^{-23/2}
   |\eta(\tau)|^{-48} 
\left [ \sum_{n,w} q^{ {{1}\over{2}}{{\alpha' p_L^2}\over{2}} }
    {\bar q}^{{{1}\over{2}} {{\alpha' p_R^2}\over{2}}} +
\left ( | \Theta_3 \Theta_4 | + | \Theta_2 \Theta_3 | + | \Theta_2 \Theta_4 |
\right ) \right ] .
\label{eq:freeee}
\end{equation}
Given the relations in Eq.\ (\ref{eq:thermodl}), for the entropy
we have the result: 
\begin{eqnarray}
S (\beta) =&& 
\half \int_{\cal F}
{{|d\tau|^2}\over{4\pi\tau_2^2}} (2\pi\tau_2)^{-23/2}
   |\eta(\tau)|^{-48} 
\sum_{n,w} \left [ 
   1 + 4 \pi \tau_2 ( - {{n^2}\over{x^2}} + {{w^2 x^2}\over{4}} ) 
\right ] 
q^{ {{1}\over{2}}{{\alpha' p_L^2}\over{2}} }
    {\bar q}^{{{1}\over{2}} {{\alpha' p_R^2}\over{2}}} \cr
\nonumber
&& \quad +
\half \int_{\cal F}
{{|d\tau|^2}\over{4\pi\tau_2^2}} (2\pi\tau_2)^{-23/2}
   |\eta(\tau)|^{-48} 
\left [ | \Theta_3 \Theta_4 | + 
| \Theta_2 \Theta_3 | + | \Theta_2 \Theta_4 |
\right ] .
\label{eq:entropy}
\end{eqnarray}
Can we give a more intuitive interpretation for the fixed point 
entropy $S_0$? Since it arises from string states localized at 
the fixed points of the orbifold, this has a plausible microscopic 
interpretation. Using the relation, $S_0$ $=$ $k_B$ 
$ {\rm ln}$ $\Omega_0$, we interpret $\Omega_0$ as the 
thermodynamic probability 
associated to the orbifold singularity in Euclidean time. 
Fortunately, this somewhat unphysical fixed point entropy will
be absent when we carry out an analogous analysis for the 
self-dual heterotic string gas \cite{fermi}. 

\section{\bf Conclusions}

This work is a satisfying resolution to two puzzles raised in 
a seminal work on string thermodynamics \cite{aw}. The first 
is the clash between the 
back-of-the-envelope argument for a Hagedorn phase transition with
an exponentially diverging free energy, 
faced with the absence of a demonstrable ultraviolet divergence in 
the one-loop effective potential or Helmholtz free energy 
\cite{polchinskibook}. The second is 
the clash of physical intuition 
with the notion of a thermal duality invariant free energy--- 
dismissed as nonsense in ref.\ \cite{aw}. We have shown that 
thermal duality follows naturally as a consequence of modular 
invariance in a closed string theory. Moreover, it leads to sensible 
results for the free string ensemble in perfect accord with physical 
intuition: the thermal duality invariant object is the 
intensive generating functional for connected one-loop vacuum 
string graphs, ${\rm ln}$ ${\rm Z}(\beta)$.
The Helmholtz and Gibbs free energies are {\em not} thermal duality
invariant, and the internal energy of the free string gas is
a monotonically increasing function of temperature.
Furthermore, as a consequence of modular invariance, in a 
tachyon-free closed string theory, the free energy of the free 
string gas does not exhibit a Hagedorn divergence \cite{fermi}.

\vskip 0.1in
We have pointed out several errors in the standard wisdom
about the Hagedorn transition in the previous literature 
on this subject 
\cite{poltorus,bowick,mclain,sathia,tan,aw,polchinskibook}.
The winding number one instability is not a signature for a
Hagedorn divergence in the free energy; the free energy of the
bosonic string gas is already divergent below $T_H$ due to the 
presence of 
{\em low temperature} tachyonic modes in the pure momentum sector.
The oft-used mapping of fundamental domain to strip in the
Polyakov path integral confuses a divergence whose physical
origin is infrared with what appears to be a divergence of
ultraviolet origin. This is permissible as a consequence of 
modular invariance \cite{polchinskibook}.
However, in the tachyon-free heterotic string
gas, it is easy to demonstrate that there is no divergence 
in the free energy at all temperatures starting from zero
\cite{fermi}.
Finally, in section 4, we have exhibited the presence of
the Kosterlitz-Thouless continuous phase transition at the
Hagedorn point. In the self-dual bosonic and heterotic 
string gases, this is a self-duality transition. In the 
type I open and closed string gas, it manifests itself as
the transition to a high temperature long string phase
as anticipated in previous works \cite{longs,sussk,polchinskibook,hp}. 
The free energy and all of its derivatives are continuous
at the critical point in each case, bosonic, heterotic, or
type I. In the tachyon-free heterotic and type I string gases, 
they are also finite and normalizable functions \cite{us,fermi}.
The peculiarities of the type II string gases are described in 
\cite{fermi}.

\vskip 0.1in
Our results give a clear analysis of the thermodynamics of the
free closed bosonic string gas within the framework of the canonical ensemble.
While many of the interesting physics applications of string 
thermodynamics await the ability to pose questions at finite 
string coupling, ideally in a nonperturbative framework, it is 
essential to have a reliable and self-consistent framework that
describes the perturbative weak coupling limit.
Of profound interest is a better understanding of the thermal
duality transition in supersymmetric string theories, and the 
spontaneous breaking of thermal duality in the 
strongly coupled heterotic string at low temperature 
\cite{us}. These are clarified in \cite{fermi}.
Fermionic string theories are replete with the possibility
of infrared instabilities at low temperature in the absence of
spacetime supersymmetry. This phenomenon is easy to exhibit 
\cite{bowick,aw,us} and its resolution brings in new concepts 
discussed in \cite{fermi}.

\vskip 0.3in
\noindent{\bf Acknowledgments}

\vskip 0.1in
\noindent 
Many of the arguments in our work can be found in preliminary form in 
the thought provoking monograph \cite{polchinskibook,poltorus}, for 
which the author is gratefully acknowledged. I am extremely grateful 
to Hikaru Kawai for insightful discussions leading to the clarified 
understanding 
of string thermodynamics given both here, and in \cite{fermi} 
(hep-th/0208112), which should be read in conjunction with this paper. 
I thank C.\ Thorn for lively email exchange that drew my 
attention to the necessity for an explanation of the absence of a 
Hagedorn phase transition in string theory. I also thank L.\ Dixon,
L.\ Dolan, D.\ 
Lowe, A.\ Shapere, and C.\ Vafa for their questions. This research was 
supported in part by the award of grant NSF-PHY-9722394 by the National 
Science Foundation under the auspices of the Career program. 

\vskip 0.5in 
\noindent{\bf Note Added (Sep 2005):} I should first correct an error, 
corrected in my papers since Aug 2004. The pressure of the string ensemble 
is non-vanishing, and
equal to the negative of the vacuum energy density. I still find the 
observation that field
theory does not distinguish between circle or orbifold topology for Euclidean time,
whereas string theory does, insightful. Notice, from the discussion
in Appendix B, that the zero temperature limit of the expression for the string 
free energy computed with the orbifold ansatze would display extraneous fixed point contributions 
absent in the original, zero temperature, vacuum energy \cite{poltorus}. 
The reason is that these extraneous terms
lack any dependence on $\beta$, as apparent in Eq.\  (\ref{eq:orbis}). Thus, 
the analysis in this paper establishes the absence of an ambiguity 
in the Euclidean time formalism for 
{\em string} statistical mechanics: the finite temperature
quantization corresponds to the background $R^{d-1}$$\times$$S^1$,
where $d$ is the critical dimension of the string theory, and $\beta$ is
the circumference of the circle. The use of the
world \lq\lq holography" to describe the $T^2$ growth of the string
free energy at high temperatures has puzzled some readers. I simply
invoked the term in order to stress the drastic thinning of high temperature
degrees of freedom in string theory
compared with the $T^d$ growth of the free energy in a $d$-dimensional field
theory. Finally, the term \lq\lq infrared instability" as used in conjunction with fermionic
string theories simply refers to the fact that the spontaneous breaking of supersymmetry
at finite temperature, consistent with modular invariance, in the type II 
superstring theories immediately leads
to tachyonic modes in the thermal spectrum. The stable endpoint of worldsheet
RG flow is the zero
temperature, 10D N=2 supersymmetric vacuum. 

\vskip 0.7in
\noindent{\large{\bf A: Absence of the Hagedorn Phase Transition}}

\vskip 0.1in
The thermal mass spectrum of the free closed bosonic string gas
contains tachyonic physical states at all temperatures starting
from zero. Each is an instability of the thermal ensemble and,
as a consequence, the duality relations and the expressions 
derived for the thermodynamic potentials in section 4 are strictly 
formal statements for the bosonic string; an illustration of 
the thermodynamics of a self-dual gas of free 
closed strings. This analysis is, however, easily repeated for the 
tachyon-free, and infrared finite, self-dual gas of free 
heterotic strings \cite{fermi}. 

\vskip 0.1in
Recall the mass formula for physical states in the untwisted
sector of the orbifold:
\begin{equation}
({\rm mass})^2 =  {{2}\over{\alpha^{\prime } }}
    ( -2 + \half {\bf l}_L^2 + 
\half {\bf l}_R^2 + N_0 + {\bar N}_0  + 
 N_{\mu} + {\bar N}_{\mu} ) \quad ,
\label{eq:masses}
\end{equation}
where we are required to project onto the subspace of states that are
invariant under the reflection: state with $N_0$ $+$ ${\bar{N}}_0$ equal to
an even integer, and lattice vectors invariant under the reflection,
$(n,w)$ $\leftrightarrow$ $(-n,-w)$.
The zero point energy in the twisted sectors differs by $+{{1}\over{16}}$,
and the oscillator moding will be half-integral. 
The level-matching condition for physical states takes the form: 
\begin{equation}
2nw + N_0 - {\bar{N}}_0 + N_{\mu}-{\bar{N}}_{\mu}=0 , \quad
\mu=1, ~ \cdots, ~ 25. 
\label{eq:lm}
\end{equation}
Both the untwisted and twisted sectors of the ${\rm Z}_2$ 
orbifold contain potentially tachyonic thermal modes.
Note that the leading tachyonic instability always occurs in the
untwisted sector since it has lower zero point energy. 
The untwisted and twisted sector tachyons with $n$ $=$ $w$ $=$ $0$
have mass squared
$ -4/\alpha^{\prime }$ and $-15/4\alpha^{\prime}$, respectively. 
Notice that these states do {\em not} contribute to the 
internal energy as can be seen in Eq.\ (\ref{eq:term}),
but are present in the Helmholtz free energy and entropy.

\vskip 0.1in
The thermal modes in the untwisted sector with 
$N_0$ $=$ ${\bar{N}}_0$ $=$ $0$, and either $n$ or $w$ non-zero, 
are potentially tachyonic over some part of the temperature 
range. They contribute in both the Helmholtz function, as well as
in the internal energy of the free string gas. 
Specifically, each momentum mode 
$(\pm n, 0)$ is tachyonic {\em upto} some critical temperature, 
$T_n$ $=$ $2/\pi n \alpha^{\prime 1/2}$, 
after which it becomes stable. Conversely, each
winding mode $(0,\pm w)$ becomes tachyonic {\em beyond} 
some critical temperature $T_w$ $=$ $w/2\pi \alpha^{\prime 1/2}$. 
Either satisfies level matching and belongs in the physical state
spectrum; requiring invariance under the Z$_2$ twist in Euclidean
time implies that we restrict to the symmetric linear combination 
of the $(+n,0)$ and $(-n,0)$ modes,
and likewise for the pure winding modes. The masses of the 
potentially tachyonic physical states are:
\begin{equation}
({\rm mass})^2_{\rm n} 
= {{2}\over{\alpha^{\prime}}} 
 \left ( -2 + {{\pi^2 n^2 \alpha^{\prime}}\over{2 \beta^2}} \right ) ,
\quad 
({\rm mass})^2_{\rm w} = {{2}\over{\alpha^{\prime}}}
 \left ( -2 + {{w^2 \beta^2}\over{ 2 \pi^2 \alpha^{\prime} }} \right ) 
 \quad . 
\label{eq:tachy}
\end{equation}
Notice that both the $(\pm 1, 0)$ and $(0,\pm 1)$ states are tachyonic
at the self-dual point, while the $(\pm 2, 0)$ and $(0,\pm 2)$ modes 
are marginally relevant.

\vskip 0.1in
The asymptotic density of string states in the zero temperature 
spectrum is known to grow exponentially as a function of mass in
the spacetime ultraviolet regime.
We will approximate the mass degeneracies, $g_N(m_N)$, 
at the $N$th mass level in the string spectrum with $N$ $>>$ $1$,
by a smooth mass density function $g(m)$
\cite{hagedorn,huangw,hr,carlitz,bowick}. 
As in the zero temperature spectrum, the asymptotic form of the density of 
states function for highly excited thermal states exhibits an exponential 
divergence, characterized once again by the Hagedorn temperature of the
bosonic string. The point is that the asymptotics of the partition function 
is identical at zero, or at finite, temperature since the asymptotics is 
characterized by the zero point energy of the 
untwisted sector of the critical string. The asymptotics in the twisted 
sector differs but the divergence is sub-leading, since the twisted sector 
has higher zero point 
energy. Consider the asymptotic form of the mass formula 
for highly excited thermal states in the untwisted sector of the orbifold:
\begin{equation}
({\rm mass})^2  \simeq  {{4}\over{\alpha^{\prime } }} N_{\rm tot} , 
\quad N_{0} = {\bar N}_{\rm 0}, \quad N_{\mu} = {\bar N}_{\rm \mu} \quad .
\label{eq:tw}
\end{equation}
We have dropped the temperature dependent zero mode
contributions in comparison to $N_0$, for large $N_0$.
The asymptotic growth of Hardy and Ramanujan's
partition sum is well-known to be exponential \cite{hr,huangw}:
\begin{equation}
\prod_{k=1}^{\infty} (1 - x^k)^{-d} \equiv \sum_{N=1}^{\infty} p(N) x^N
\quad \quad  \lim_{N \to \infty} p(N) = 
  N^{-(d+3)/4} e^{ 2 \pi(d/6)^{1/2} N^{1/2}}
\quad .
\label{eq:hr}
\end{equation}
At high levels, the masses in the free string spectrum
approach a continuum, and the approximation of a smooth number density
as a function of mass, $g(m)$, is usually made in the string literature
\cite{huangw,carlitz,bowick,tan,aw,polchinskibook}. 
In fact, by substitution from above, the asymptotic mass density, 
$g(m)$, for states of mass, $m$, in 
the untwisted sector of the free bosonic
string gas is found to take the form \cite{huangw}:
\begin{equation}
g(m) \simeq  e^{4 \pi \alpha^{\prime 1/2} m }  \quad .
\label{eq:massden}
\end{equation}

\vskip 0.1in
Exponential growth in the mass density of a {\em particle}-like
thermodynamic ensemble usually indicates the existence of a 
characteristic exponential divergence in the free energy, as 
first noted by Hagedorn \cite{hagedorn}. The argument is simple.
Consider a statistical ensemble of particles
described by a Hamiltonian, $H$, with
eigenspectrum described by an exponentially growing
mass density, $g(m)$ $=$ $e^{ b m}$, with $b$ some characteristic
length scale. The Helmholtz free energy of the particle
ensemble can be written as \cite{hagedorn,huangw}:
\begin{equation}
F (\beta ) =
V \int_{m_0}^{\infty} dm g(m) e^{- \beta H}
= V \int_{m_0}^{\infty} dm e^{- (\beta - b) m }
  \quad ,
\label{eq:ht}
\end{equation}
where $m_0$ is an infrared cutoff on the mass spectrum 
of the statistical ensemble below which the exponential form
assumed for $g(m)$ may not hold. The free energy is seen to 
diverge exponentially for temperatures $\beta^{-1}$ $>$ $1/b$, and the 
critical value,
$1/b$, is known as the Hagedorn transition temperature. Thus,
exponential growth in the asymptotic density of particle 
states suggests a phase transition, or limiting temperature,
$1/\beta_H$, beyond which the free energy diverges 
exponentially.

\vskip 0.1in
Notice that this argument does {\em not} hold for the gas of free closed 
strings. Despite exponential growth in the asymptotic density of
states function in the free string spectrum, there is no 
corresponding exponential divergence in the Helmholtz free energy. 
The reason, as explained in the introduction, is that the 
one-loop contribution to the
free energy in string theory is
given by an integral over world-sheet moduli of a modular
invariant function, ${\rm Z}(\tau,{\bar{\tau}})$. 
As a consequence, the integration
over the moduli, $\tau$, is restricted to the fundamental 
domain of the one-loop modular group, which does not include the region 
of $\tau$ space for which the density of states 
function diverges exponentially. Hence, there is no exponentially 
diverging free energy at one-loop and, consequently, no Hagedorn 
phase transition in string theory. An analogous argument can, in
fact, be constructed at any loop in the perturbative expansion for
string theory \cite{rg}. 

\vskip 0.1in
Previous authors have 
mistakenly referred to the exponential divergence due to the 
onset of the tachyonic instability in the winding number one 
thermal mode
as a \lq\lq Hagedorn phase transition". As explained in the 
introduction, this is incorrect. The misleading exponential 
divergence in \cite{poltorus,mclain,longs,sathia,tan,aw,kani,hp} of 
apparently ultraviolet origin, arises as a consequence of
a Poisson resummation of the direct exponential divergence in the
partition function due to the leading tachyonic winding mode in
the spectrum. We emphasize that the mistaken identification of 
the w=1 tachyonic instability as a signal for a Hagedorn phase 
transition, ignores the proliferation of low temperature 
tachyonic modes with $n$$>$$0$, $N_0$$=$${\bar N}_0$$=$$0$,
that are also in the physical state spectrum.
The free energy is already divergent at temperatures much 
below $T_H$. Thus, the winding number one tachyonic instability 
cannot be interpreted as a Hagedorn phase transition in the free 
string gas. To verify our claim, in \cite{fermi} we compute 
the free energy for a free heterotic string gas without 
tachyons in the thermal spectrum. The free energy is found to be 
finite at all temperatures starting from zero including the 
Hagedorn temperature.

\vskip 0.5in
\noindent{{\large{\bf B:  Fixed Point Contributions to the Entropy}}

\vskip 0.1in
As mentioned in the introduction, the expressions we will
derive for both the Helmholtz free energy and the entropy 
function will contain temperature independent contributions
from string states localized at the fixed points of the 
orbifold. Since it is of some interest to understand their
explicit form, in this subsection we give a brief review of 
the relationship between ${\rm Z}_2$ orbifold and circle 
compactifications at generic radius \cite{polchinskibook}. 
Recall the precise
equivalence of the partition functions of a free 2d boson 
compactified on a circle of radius 
$2\alpha^{\prime 1/2}$ or on an interval of length 
$\pi \alpha^{\prime 1/2}$, i.e.,  
the partition function on the ${\rm Z}_2$ orbifold of the
circle of radius $\alpha^{\prime 1/2}$. This marks the 
Kosterlitz-Thouless
point: the intersection of the circle and orbifold fixed 
lines in the moduli space of c=1 conformal field theories
\cite{ginsparg,polchinskibook}.
Let us review the derivation of the fixed point contributions
following \cite{ginsparg,polchinskibook}.

\vskip 0.1in
The untwisted sector of the orbifold contains the subset of
states in the circle compactification invariant under 
the reflection, $X^0$$\to$$-X^0$, with fixed points at 
$X^0$$=$$0$, $\pi r_{\rm circ.}$. 
The untwisted sector states have even
oscillator number: $N^0$$+$${\bar{N}}^0$, 
and their momentum vectors lie in the subspace
of lattice vectors invariant under: $(n,w)$$\to$$(-n,-w)$. 
Thus, for the untwisted sector, we have:
\begin{equation}
(q{\bar{q}})^{-1/24} {\rm tr}_U {{{\bf 1} + {\bf R} }\over{2}} 
 q^{L_0} {\bar{q}}^{{\bar{L}}_0} = \half \left (
Z_{\rm circ.}(x) + 
(q {\bar{q}})^{1/24} \prod_{m=1}^{\infty} |1+q^m|^{-2} 
\right )
\quad .
\label{eq:untwist}
\end{equation} 
This expression is not modular invariant due to the presence of the
second term: the projection to states invariant under the reflection
spoils the Lorentzian self-duality property of the lattice vectors 
summed over in Eq.\ (\ref{eq:untwist}). A modular invariant spectrum 
of free strings living on the interval requires inclusion
of additional twisted sectors in the path integral,
where $X^0$ satisfies the boundary condition 
$X^0(\sigma^2 + 2\pi) = - X^0 (\sigma^1)$. This implies the absence
of the zero mode, $n$$=$$w$$=$$0$, and the twisted 
strings are therefore localized at one of two fixed points,
$X^0$$=$$0$, $\pi r_{\rm circ.}$. 
The anti-periodicity in the boundary
condition implies a half-integer mode expansion.
The result for the sum of the twisted sectors
is \cite{polchinskibook}:
\begin{equation}
(q{\bar{q}})^{1/48} {\rm tr}_{\em T} {{{\bf 1} + {\bf R} }\over{2}} 
 q^{L_0} {\bar{q}}^{{\bar{L}}_0} = \half \left (
(q {\bar{q}})^{1/48} \prod_{m=1}^{\infty} |1-q^{m-1/2}|^{-2} 
+
(q {\bar{q}})^{1/48} \prod_{m=1}^{\infty} |1+q^{m-1/2}|^{-2} 
\right )
\quad .
\label{eq:twist}
\end{equation} 
Combining Eqs.\ (\ref{eq:untwist}) and (\ref{eq:twist}), the
final result can be expressed in the form given in Eq.\ (\ref{eq:boso}) 
\cite{ginsparg} by using the Jacobi triple product identity.
The $\Theta_i(0,\tau)$ are the Jacobi theta functions.  
Notice the manifest modular invariance of the expression in
Eq.\ (\ref{eq:poiss}): under $\tau $$\to$$\tau$$+$$1$, 
$\Theta_2$$\to$$-\Theta_2$, $\Theta_3$$\to$$\Theta_4$, and 
vice versa. Under $\tau$$\to$$-1/\tau$, $\Theta_3$$\to$$(-i\tau)^{1/2}
\Theta_3$,  
$\Theta_4$$\to$$(-i\tau_2)^{1/2} \Theta_4$. 
The absolute magnitudes remove the phases.
Thus, the partition function, ${\rm Z}(\tau ,
{\bar{\tau}})$, for the $24$ transverse bosonic degrees 
of freedom in the critical bosonic string 
can be written in the form:
\begin{equation}
(2\pi \tau_2)^{-23/2} |\eta(\tau)|^{-46} \cdot \left \{ |\eta(\tau)|^{-2}
\left [ \sum_{n,w} q^{ {{1}\over{2}}{{\alpha' p_L^2}\over{2}} }
    {\bar q}^{{{1}\over{2}} {{\alpha' p_R^2}\over{2}}} +
\left ( | \Theta_3 \Theta_4 | + | \Theta_2 \Theta_3 | + | \Theta_2 \Theta_4 | \right )
\right ] \right \} = {\rm Z}_{23} \cdot {\rm Z}_{{\rm orb.}} ,
\label{eq:part}
\end{equation}
as given in Eq.\ (\ref{eq:boso}). 
The factor inside curly brackets is the orbifold partition function,
and ${\rm Z}_{23}$ is the partition function of $23$ free bosons
with noncompact target space $R^{23}$. Each factor is separately 
modular invariant.

\vskip 0.1in
Recall that the spectrum of states in 
the partition function of a free boson are
in one-to-one correspondence with the operators in a two dimensional conformal
field theory with central charge $c$ $=$ $1$ \cite{ginsparg}. The spectrum
contains a single marginal operator, $\partial_z X^0 \partial_{\bar{z}} X^0 $,
of conformal dimension $(h,{\bar h})$$=$$(1,1)$.
Perturbation of the free Lagrangian by this operator leaves the spectrum and
quantum correlation functions of the conformal field theory unchanged, but
for a change in the radius of the target space. In thermal string
theory, this modulus is the inverse temperature, the size of the
interval in the imaginary time direction.
The $c$ $=$ $1$ conformal field theories have been classified, and 
the form of their moduli space is known \cite{ginsparg,polchinskibook}. 
The two fixed lines correspond to two-dimensional
free boson theories whose target spaces are,
respectively, a circle and an orbifold. The fixed lines intersect at the
continuum limit of the Kosterlitz-Thouless (KT) point of the X-Y model 
\cite{kosterlitz}.
The partition function on a circle of radius $r_{\rm circ}$
is related to the partition function on the Z$_2$-orbifold of 
the circle with identical radius as follows \cite{ginsparg}:
\begin{equation}
{\rm Z}_{\rm orb.} (r_{\rm circ.}) = \half \left [ {\rm Z}_{\rm circ.} (r_{\rm circ.})
+ 2 {\rm Z}_{\rm circ.} ( \alpha^{\prime 1/2} ) -
 {\rm Z}_{\rm circ.} (2 \alpha^{\prime 1/2} )
\right ]
\quad .
\label{eq:orbis}
\end{equation}
Points on the circle fixed line correspond to $S^1$ 
theories with radius $r_{\rm circ.}$, while points
on the orbifold fixed line correspond to $S^1/{\rm Z}_{2}$ theories 
with interval length $\pi r_{\rm circ.}$. 
Setting $r_{\rm circ.}$ $=$ ${{\alpha^{\prime 1/2}}\over{2}}$ in
Eq.\ (\ref{eq:orbis}), gives the
equality of orbifold and circle partition functions at the KT point of
the $c$ $=$ $1$ moduli space:
\begin{equation}
{\rm Z}_{\rm orb.} ( \alpha^{\prime 1/2} ) =
 {\rm Z}_{\rm circ.} ( {{\alpha^{\prime 1/2}}\over{2}} ) =
 {\rm Z}_{\rm circ.} ( 2\alpha^{\prime 1/2} ) \equiv {\rm Z}_{KT}
\quad .
\label{eq:orbkt}
\end{equation}
The second equality follows from thermal duality of the circle
partition function. 
Thermal duality is a simple consequence of Lorentz invariance together
with the invariance of the vacuum energy functional of a closed string
theory under T-duality transformations mapping small radii to large
radii, $r$$\to$$\alpha^{\prime}/r$ \cite{polchinskibook}.
Note that either fixed line in the $c$ $=$ $1$ moduli space displays an
$r$$\to$$r^2_c/r$ duality, with self-dual radius,
$r_c$ $=$ $ \alpha^{\prime 1/2}$. Thus, the partition function of the
KT model has two equivalent representations: as the circle theory
of radius $2\alpha^{\prime}$, or as the R-orbifold theory 
with interval length $\pi \alpha^{\prime 1/2}$.

\end{document}